\RequirePackage[l2tabu, orthodox]{nag}
\RequirePackage{snapshot}

\documentclass[10pt,twocolumn]{article}

\usepackage[dvips,letterpaper,top=0.5in, bottom=0.5in, left=0.75in, right=0.5in,includefoot,heightrounded]{geometry}

\usepackage{srcltx}

  \usepackage{lipsum}
\usepackage{amsmath}
\usepackage{amssymb,amsfonts}

\usepackage{abstract}

\usepackage{epsfig}
\usepackage[usenames,dvipsnames]{color}
\usepackage[usenames,dvipsnames]{xcolor}
\usepackage{subfigure}

\usepackage{booktabs}

\usepackage{setspace}
\usepackage{multicol}

\usepackage{cite}
\usepackage{url}
\usepackage[normalem]{ulem}

\usepackage{enumerate}

\usepackage{hyperref}

\usepackage[sc,tiny]{titlesec}
       \usepackage{mathptmx}

\newtheorem{proposition}{Proposition}
\newtheorem{definition}{Definition}
\newtheorem{theorem}{Theorem}
\newtheorem{lemma}{Lemma}
\newtheorem{corollary}{Corollary}

\def\QED{\mbox{$\square$}}
\def\proof{\noindent{\it Proof:~}}
\def\endproof{\hspace*{\fill}~\QED\par\endtrivlist\unskip}

\newcommand{\dens}{\operatorname{dens}}
\newcommand{\asymdens}{\operatorname{d}}
\newcommand{\lasymdens}{\operatorname{\underline{d}}}
\newcommand{\uasymdens}{\operatorname{\bar{d}}}
\newcommand{\logdens}{\operatorname{\ell d}}
\newcommand{\dirichletdens}{\operatorname{D}}
\newcommand{\schdens}{\operatorname{\delta}}

\newcommand{\li}{\operatorname{li}}



\title{%
On a Density for Sets of Integers
}

\author{%
R.~J.~Cintra%
\thanks{%
R.~J.~Cintra is with 
the Signal Processing Group,
Departamento de Estat\'{\i}stica, 
Universidade Federal de Pernambuco.
Email: 
\protect\url{rjdsc@de.ufpe.br}
}
\and
L.~C.~R\^ego%
\thanks{%
L.~C.~R\^ego is with 
Departamento de Estat\'{\i}stica, 
Universidade Federal de Pernambuco.
Email: 
\protect\url{leandro@de.ufpe.br}
}
\and
H. M. de Oliveira%
\thanks{%
H. M. de Olveira was with 
the Departamento de Eletr\^onica e Sistemas
Universidade Federal de Pernambuco.
Now he is with
the Signal Processing Group,
Departamento de Estat\'{\i}stica, 
Universidade Federal de Pernambuco.
Email: 
\protect\url{hmo@de.ufpe.br}
}
\and
R. M. Campello de Souza%
\thanks{%
R. M. Campello de Souza is with 
the Departamento de Eletr\^onica e Sistemas
Universidade Federal de Pernambuco.
Email:
\protect\url{ricardo@ufpe.br}
}
}

\date{}

\newcommand{\myabstract}{%
A relationship between the Riemann zeta function and a density on
integer sets is explored.
Several properties of the examined density are derived.
}

\newcommand{\mykeywords}{%
Number theory, probability theory, arithmetization
}

\begin{document}

\makeatletter
\if@twocolumn

\twocolumn[%
  \maketitle
  \begin{onecolabstract}
    \myabstract
  \end{onecolabstract}
  \begin{center}
    \medskip
    \mykeywords
  \end{center}
  \bigskip
]
\saythanks

\else

  \maketitle
  \begin{abstract}
    \myabstract
  \end{abstract}
  \begin{center}
    \medskip
    \mykeywords
  \end{center}
  \bigskip
  \onehalfspacing
\fi

\section{Introduction}

Several measures for the density of sets have been discussed in the
literature~\cite{bell2006,duncan1968,duncan1970,erdos1948,davenport1951,ahlswede1997}.
Presumably the most employed tool for evaluating the density of sets is
the asymptotic density, also referred to as natural density.
The asymptotic density is expressed by
\begin{equation}
\asymdens(A) =
\lim_{n\to\infty}
\frac{\|A \cap \{1, 2, 3, \ldots, n\}\|}{n},
\end{equation}
provided that such a limit does exist.
The symbol $\|\cdot\|$ denotes the cardinality,
and $A$ is a set of integers.
Analogously, the lower and upper asymptotic densities are defined by
\begin{align}
\lasymdens(A)
&=
\liminf_{n\to\infty}
\frac{\|A \cap \{1, 2, 3, \ldots, n\}\|}{n},
\\
\uasymdens(A)
&=
\limsup_{n\to\infty}
\frac{\|A \cap \{1, 2, 3, \ldots, n\}\|}{n},
\end{align}
The asymptotic density is said to exist
if and only if both the lower and upper asymptotic densities do exist and are equal.

Although the asymptotic density does not always exist,
the Schnirelmann density~\cite{duncan1968,duncan1970}
is always well-defined.
The Schnirelmann density is defined as
\begin{align}
\schdens(A)
=
\inf_{n\geq 1}
\frac{\|A \cap \{1, 2, 3, \ldots, n\}\|}{n}.
\end{align}
Interestingly, this density is highly sensitive to the initial
elements of sequence.
For instance, if $1 \not \in A$, then $\schdens(A) = 0$~\cite{niven1960}.

Another interesting tool is the
logarithmic density~\cite{erdos1948,davenport1951,ahlswede1997}.
Let $A = \{a_1, a_2, a_3, \ldots\}$ be a set of integers.
The logarithmic density of $A$ is given by
\begin{equation}
\logdens(A)
=
\lim_{n\to\infty}
\frac{\sum_{a_i \leq n} \frac{1}{a_i}}{\log n}.
\end{equation}

In~\cite{bell2006}, Bell and Burris
bring a good exposition on the Dirichlet density.
The Dirichlet density
is defined as the limit of
the ratio between two Dirichlet series.
Let $A\subset B$ be two sets.
The \emph{generating series} of $A$
is given by
\begin{align}
\mathbf{A}(s)
=
\sum_{n=1}^\infty
\frac{N(A,n)}{n^s},
\end{align}
where $N(A,n)$ is a counting function that returns the number of elements in $A$ of norm $n$~\cite{bell2006}.
The Dirichlet density is then expressed by
\begin{align}
\label{def.dirichlet.bell-burris}
\dirichletdens(A)
=
\lim_{s\downarrow\alpha}
\frac{\mathbf{A}(s)}{\mathbf{B}(s)},
\end{align}
where $\mathbf{B}(s)$ is the generating series of $B$
and $\alpha$ is an abscissa of convergence~\cite{bell2006}.
In~\cite{ahlswede1997}, we also find the Dirichlet density
defined as
\begin{equation}
\dirichletdens(A)
=
\lim_{s\downarrow 1}
(s-1)
\sum_{n\in A}\frac{1}{n^s},
\end{equation}
whenever the limit exists.
This density admits lower and upper versions,
simply by replacing the above limit by $\liminf$ and $\limsup$,
respectively.

The aim of this work is to investigate the properties of the Dirichlet density
as defined in Equation~\ref{def.dirichlet.bell-burris}
in the particular case where the set $B$ is the set of natural numbers.
This induces a density based on the Riemann zeta function.

\section{A Density for Sets of Integers}

In this section,
we investigate the
particular case of the Dirichlet function,
applied for
subsets of the natural numbers.
In this case,
taking into consideration the usual norm,
where the norm of any natural number is equal to its absolute value,
the counting function $N(\cdot,\cdot)$ becomes
\begin{align}
N(A,n)
=
\begin{cases}
1, & \text{if $n\in A$,} \\
0, & \text{otherwise.}
\end{cases}
\end{align}

\begin{definition}
The density $\dens$ of a subset $A\subset\mathbb{N}$
is given by
\begin{align}
\label{equation.mu.s}
\dens(A)
&\triangleq
\lim_{s\downarrow1}
\frac
{\sum_{n=1}^\infty  N(A,n)\frac{1}{n^s}}
{\sum_{n=1}^\infty  N(\mathbb{N},n)\frac{1}{n^s}}\\
&=
\lim_{s\downarrow1}
\frac{\sum_{n\in A}\frac{1}{n^s}}
{\sum_{n=1}^\infty\frac{1}{n^s}}
\\
&=
\lim_{s\downarrow1}
\frac{\sum_{n\in A}\frac{1}{n^s}}
{\zeta(s)},
\quad
s>1,
\end{align}
if the limit exists.
The quantity $\zeta(\cdot)$ denotes the Riemann zeta function~\cite{grad1965}.
\end{definition}

\begin{proposition}
The following assertions hold true:
\begin{enumerate}
\item
$\dens(\mathbb{N}) = 1$

\item
$\dens(A) \geq 0$ (nonnegativity)

\item
if $\dens(A)$ and $\dens(B)$ exist%
\footnote{For ease of exposition,
in the following results,
we assume that the densities of the relevant sets always do exist.}
and $A\cap B = \varnothing$, then $\dens(A\cup B)$ exists and is equal to
$$\dens(A)+\dens(B)\mbox{ (additivity).}$$
\end{enumerate}
\end{proposition}

\proof
Statements 1 and 2 cab be trivially checked.
The additivity property can be derived as follows:
\begin{align}
\dens(A\cup B)
&=
\lim_{s\downarrow1}
\frac{\sum_{n\in A\cup B} \frac{1}{n^s}}
{\zeta(s)}
\\
&=
\lim_{s\downarrow1}
\frac{\sum_{n\in A} \frac{1}{n^s}}
{\zeta(s)}
+
\lim_{s\downarrow1}
\frac{\sum_{n\in B} \frac{1}{n^s}}
{\zeta(s)}
\\
&=
\dens(A) + \dens(B).
\end{align}
\endproof

\begin{corollary}
The density of the null set is zero.
\end{corollary}
\proof
In fact, $\dens(\varnothing) = 0$,
since
$1 = \dens(\mathbb{N}) = \dens(\varnothing \cup \mathbb{N}) = \dens(\varnothing)+\dens(\mathbb{N}) = \dens(\varnothing) + 1$.
\endproof

\begin{corollary}
$\dens(B-A) = \dens(B) - \dens(A\cap B)$.
\end{corollary}
\proof
The proof is straightforward:
\begin{align}
\dens(B)
&=
\dens((A\cap B) \cup (A^c\cap B))
\\
&=
\dens(A\cap B) + \dens(A^c\cap B)
\\
&=
\dens(A\cap B) + \dens(B-A),
\end{align}
since $A\cap B$ and $A^c\cap B$ are disjoint.
\endproof

\begin{corollary}
For every $A$,
$\dens(A^c) = 1 - \dens(A)$,
where $A^c$ is the complement of $A$.
\end{corollary}
\proof
$\dens(A^c)
=
\dens(\mathbb{N} - A)
=
\dens(\mathbb{N})-\dens(\mathbb{N}\cap A)
=
1 - \dens(A)$.

\begin{proposition}[Monotonicity]
The discussed density is a monotone function, i.e.,
if $A\subset B$, then $\dens(A) \leq \dens(B)$.
\end{proposition}
\proof
We have that
\begin{align}
\dens(B)
&=
\lim_{s\downarrow1}
\frac{\sum_{n\in B}\frac{1}{n^s}}
{\zeta(s)}
\\
&=
\lim_{s\downarrow1}
\frac{\sum_{n\in A}\frac{1}{n^s}}
{\zeta(s)}
+
\lim_{s\downarrow1}
\frac{\sum_{n\in B-A}\frac{1}{n^s}}
{\zeta(s)}
\\
&\geq
\lim_{s\downarrow1}
\frac{\sum_{n\in A}\frac{1}{n^s}}
{\zeta(s)}
\\
&=
\dens(A).
\end{align}
\endproof

\begin{proposition}[Finite Sets]
Every finite subset of $\mathbb{N}$
has density zero.
\end{proposition}
\proof
Let $A$ be a finite set.
For $s>1$, it yields
\begin{equation}
\sum_{n\in A} \frac{1}{n^s} \leq \sum_{n\in A} \frac{1}{n} = \mathrm{constant} < \infty.
\end{equation}
Thus,
\begin{equation}
\dens(A) =
\lim_{s\downarrow1}
\frac{\sum_{n\in A}\frac{1}{n^s}}
{\zeta(s)}
\leq
\lim_{s\downarrow1}
\frac{\mathrm{constant}}{\zeta(s)}
=
0.
\end{equation}
\endproof
\noindent
As a consequence,
sets $A\subset\mathbb{N}$ of nonzero density must be infinite.

\begin{corollary}
The density of a singleton is zero.
\end{corollary}

\begin{proposition}[Union]
\label{proposition.union}
Let $A$ and $B$ be two sets of integers.
Then
the density of $A \cup B$ is given by
\begin{equation}
\dens(A \cup B) = \dens(A) + \dens(B) - \dens(A \cap B).
\end{equation}
\end{proposition}
\proof
Observe that $A \cup B = A \cup (B-A)$, and $A\cap (B-A) = \varnothing$.
Then, it follows directly from the properties of the density measure that
\begin{align}
\dens(A \cup B)&=\dens(A \cup (B-A)) \\
&=\dens(A) + \dens(B-A) \\
&=\dens(A) + \dens(B) - \dens(A \cap B).
\end{align}
\endproof

\begin{proposition}[Heavy Tail]
Let $A = \{a_1, a_2, \ldots, a_{N-1}, a_N, a_{N+1}, \ldots \}$.
If $A_1 = \{a_1, a_2, \ldots, a_{N-1}\}$
and
$A_2 = \{a_N, a_{N+1}, \ldots \}$,
then
\begin{align}
\dens(A) = \dens(A_2).
\end{align}
\end{proposition}
\proof
Observe that $A = A_1 \cup A_2$ and  $A_1$ and $A_2$ are disjoint.
Therefore, $\dens(A) = \dens(A_1) + \dens(A_2)$.
Since $A_1$ is a finite set, $\dens(A_1)=0$.
\endproof

\bigskip

Now consider the following operation $m \otimes A \triangleq \{ m a \ |\ a\in A, m\in\mathbb{N}\}$.
This can be interpreted as a dilation operation on the elements of $A$.

In~\cite{erdos1994,davenport1951,erdos1948},
Erd\"os \emph{et al.}
examined the density of the set of multiples $m \otimes A$,
showing the existence of a logarithmic density equal to its lower asymptotic density.
Herein we investigate further this matter, evaluating the density of
sets of multiples.

\begin{proposition}[Dilation]
\label{proposition.scaling}
Let $A$ be a set, such as $\dens(A)>0$.
Then
\begin{equation}
\dens(m \otimes A) = \frac{1}{m}\dens(A).
\end{equation}
\end{proposition}
\proof
This result follows directly from the definition of the discussed density:
\begin{align}
\dens(m \otimes A)
&=
\lim_{s\downarrow1}
\frac{\sum_{n\in A} \frac{1}{(mn)^s}}
{\zeta(s)}
\\
&=
\lim_{s\downarrow1}
\frac{\frac{1}{m^s}\sum_{n\in A} \frac{1}{n^s}}
{\zeta(s)}
\\
&=
\frac{1}{m}
\dens(A).
\end{align}
\endproof

Let $A\oplus m \triangleq \{ a + m \ |\ a\in A, m\in\mathbb{N}\}$.
This process is called a translation of $A$ by $m$ units~\cite[p.49]{rudin1970}.
Our aim is to show that the discussed density is translation invariant, i.e.,
$\dens(A \oplus m) = \dens(A)$, $m>0$.
Before that we need the following lemma.

\begin{lemma}[Unitary Translation]
Let $A$ be a set, such as $\dens(A)>0$.
Then
\begin{equation}
\dens(A \oplus 1) = \dens(A).
\end{equation}
\end{lemma}
\proof
Let $A = \{a_1, a_2, \ldots, a_{N-1}, a_N, a_{N+1}, \ldots \}$.
We can split $A$ into two disjoint sets as shown below:
\begin{align}
A
& =
\{a_1, a_2, \ldots, a_{N-1}, a_N, a_{N+1}, \ldots \}
\\
&=
\{a_1, a_2, \ldots, a_{N-1}\}
\cup
\{a_N, a_{N+1}, \ldots \}
\\
&=
A_1^N \cup A_2^N,
\end{align}
where $A_1^N$ is a finite set and $A_2^N$ is a `tail' set starting at the element $a_N$.
By the Heavy Tail property, we have that $\forall N$
$\dens(A) = \dens(A_2^N)$.
For the same reason, $\forall N$, $\dens(A \oplus 1) = \dens(A_2^N \oplus 1)$.

Note that, for $s>1$, if $kn \geq n+1$, then the following inequalities hold:
\begin{align}
\frac{\sum_{n\in A_2^N} \frac{1}{(kn)^s}}
{\zeta(s)}
\leq
\frac{\sum_{n\in A_2^N} \frac{1}{(n+1)^s}}
{\zeta(s)}
\leq
\frac{\sum_{n\in A_2^N} \frac{1}{n^s}}
{\zeta(s)}.
\end{align}
In other words,
for the above inequality to be valid,
we must have
\begin{align}
k \geq \frac{n+1}{n}, \qquad \forall n \in A_2^N.
\end{align}
For instance, let $k = (a_N+1)/a_N$.
Consequently, we establish that
\begin{align}
\frac{\sum_{n\in A_2^N} \frac{1}{( \frac{a_N+1}{a_N}  n)^s}}
{\zeta(s)}
\leq
\frac{\sum_{n\in A_2^N} \frac{1}{(n+1)^s}}
{\zeta(s)}
\leq
\frac{\sum_{n\in A_2^N} \frac{1}{n^s}}
{\zeta(s)}.
\end{align}
Taking the limit as $s\downarrow1$, the above upper bound
becomes
\begin{align}
\lim_{s\downarrow1}
\frac{\sum_{n\in A_2^N} \frac{1}{n^s}}
{\zeta(s)}
=
\dens(A_2^N)=\dens(A).
\end{align}
Now examining the lower bound and using the dilation property, we have that
\begin{align}
&\lim_{s\downarrow1}
\frac{\sum_{n\in A_2^N} \frac{1}{( \frac{a_N+1}{a_N}  n)^s}}
{\zeta(s)}\\
&=
\frac{a_N}{a_N+1}
\dens(A_2^N)\\
&=
\frac{a_N}{a_N+1}
\dens(A).
\end{align}
Since $\forall \epsilon > 0$, $\exists N$ such that
\begin{equation}
\frac{a_N}{a_N+1} \dens(A) > \dens(A) - \epsilon,
\end{equation}
and as
\begin{align}
\lim_{s\downarrow1}
\frac{\sum_{n\in A_2^N} \frac{1}{(n+1)^s}}
{\zeta(s)}
=
\dens(A_2^N\oplus 1)=\dens(A\oplus 1),
\end{align}
it follows that $\forall \epsilon>0$ we have that
\begin{align}
\dens(A) - \epsilon
\leq
\dens(A\oplus 1)
\leq \dens(A).
\end{align}

Therefore, letting $\epsilon\to0$, it follows that
\begin{align}
\dens(A\oplus 1)
=\dens(A).
\end{align}
\endproof

\begin{proposition}[Translation Invariance]
Let $A$ be a set, such as $\dens(A)>0$.
Then
\begin{equation}
\dens(A \oplus m) = \dens(A),
\end{equation}
where $m$ is a positive integer.
\end{proposition}
\proof
The proof follows by finite induction.
We have already proven that $\dens(A \oplus 1)=\dens(A)$.
Therefore, we have that
\begin{align}
&\dens(A \oplus (m+1)) = \dens((A \oplus m) \oplus 1)\\
&=
\dens(A \oplus m)=\dens(A).
\end{align}
\endproof

One possible application for a density on a set of natural numbers
$A$ is to interpret it as the chance of choosing a natural number in
$A$ when all natural numbers are equally likely to be chosen.
Interestingly, as we show next the above measure of uncertainty does
not obey all the axioms of Kolmogorov since it is not
$\sigma$-additive.
Additionally, we show that it is impossible to define
a finite $\sigma$-additive translation invariant measure on
$(\mathbb{N},2^{\mathbb{N}})$.
This result emphasizes an important
point that there are reasonable measures of uncertainty that do not
satisfy the formal standard definition of a probability measure.

\begin{theorem}
There is no $\sigma$-additive measure defined on the
measurable space $(\mathbb{N},2^{\mathbb{N}})$ such that:
\begin{enumerate}
\item $0<\mu(\mathbb{N})<\infty$; and
\item $\mu$ is translation invariant.
\end{enumerate}
\end{theorem}
\proof
Suppose that $\mu$ is translation invariant.
Then
every singleton set must have the same measure.
Let
$\omega_1<\omega_2$ be any natural numbers. Since
$\{\omega_1\}\oplus(\omega_2-\omega_1)=\{\omega_2\}$, we have
$$\mu(\{\omega_2\})=\mu(\{\omega_1\}\oplus(\omega_2-\omega_1))=\mu(\{\omega_1\}),$$
where the last equality follows from translation invariance. Let
$c=\mu(\{1\})$.
If $c=0$, then $\mu(\mathbb{N})\ne
0=\sum_{i=1}^{\infty}\mu(\{i\})$.
Thus $\mu$ is not $\sigma$-additive.
If $c>0$, then $\mu(\mathbb{N})<\infty=\sum_{i=1}^{\infty}\mu(\{i\})$,
which also implies that $\mu$ is not $\sigma$-additive.
\endproof

\begin{proposition}[Criterion for Zero Density]

Let $A=\{a_1, a_2, \ldots, a_n, a_{n+1}, \ldots\}$.
If
\ $\lim_{s\downarrow1}\sum_{n=1}^\infty\frac{1}{a_{n}^s}$ converges,
then
$\dens(A) = 0$.
\end{proposition}
\proof
It follows directly from the definition of $\dens(A)$.
\endproof

\begin{definition}[Sparse Set]
A zero density set is said to be a sparse set.
\end{definition}

\noindent
As matter of fact, any criterion that ensure the convergence of
$\lim_{s\downarrow1}\sum_{n=1}^\infty\frac{1}{a_{n}^s}$
can be taken into consideration.
In the next result, we utilize the Ratio Test for convergence of series.
\begin{corollary}
Let $A=\{a_1, a_2, \ldots, a_n, a_{n+1}, \ldots\}$.
If
\ $\lim_{n\to\infty}\frac{a_n}{a_{n+1}} < 1$,
then
$\dens(A) = 0$.
\end{corollary}
\proof
Observe that for $s>1$
\begin{align}
\sum_{n\in A}\frac{1}{n^s}
<
\sum_{n\in A}\frac{1}{n}
=
\sum_{n=1}^\infty\frac{1}{a_n}
.
\end{align}
According to the Ratio Test~\cite[p.68]{abbott2001},
the series on the right-hand side of the above inequality
converges whenever
\begin{align}
\lim_{n\to\infty}
\frac{\frac{1}{a_{n+1}}}{\frac{1}{a_n}}
=
\lim_{n\to\infty}
\frac{a_n}{a_{n+1}}
<
1.
\end{align}
Thus, by series dominance,
it follows that
$\sum_{n\in A}\frac{1}{n^s}$ also converges.
And finally, this implies that
the density of $A$ must be zero.
\endproof

\begin{lemma}[Powers of the Set Elements]
Let $A = \{a_1, a_2, \ldots\}$ and $m>1$ be an integer.
The density of $A^m \triangleq \{a_1^m, a_2^m, \ldots\}$
is zero.
\end{lemma}
\proof
Notice that, for $s > 1$ and $m>1$,
\begin{align}
\sum_{n=1}^\infty
\frac{1}{(a_n^m)^s}
<
\sum_{n = 1}^\infty
\frac{1}{a_n^m}
<
\sum_{n = 1}^\infty
\frac{1}{n^m}
<
\infty
\end{align}
Thus, it follows from the preceding discussion that $\dens(A) = 0$.
\endproof

\begin{corollary}
The density of the set of perfect squares is zero.
\end{corollary}

\begin{corollary}[Geometric Progressions]
Let $G=\{ar, ar^2, ar^3, \ldots\}$, where $a$ and $r>1$ are positive integers.
Then
\begin{equation}
\dens(G) = 0.
\end{equation}
\end{corollary}
\proof
Notice that, for $r > 1$,
\begin{align}
\sum_{n=1}^\infty
\frac{1}{(ar^n)^s}
=\frac{1}{a^s}
\sum_{n = 1}^\infty
\frac{1}{r^{ns}}
<\frac{1}{a^s}
\sum_{n = 1}^\infty
\left(\frac{1}{r}\right)^n
<
\infty
\end{align}
Thus, it also follows from the preceding discussion that $\dens(G) = 0$.
\endproof

Let $M_p = \{p, 2p, 3p, \ldots\}$ be an arithmetic progression, where $p\in\mathbb{N}$.
For example,
$M_2 = \{2, 4, 6, \ldots\}$,
$M_5 = \{5, 10, 15, \ldots\}$, etc.
Regarding the cardinality of these sets,
we have that $\|M_p\| = \aleph_0$.

\begin{proposition}[Arithmetic Progressions]
For a fixed integer $p$, the density of the set $M_p$ is given by
\begin{equation}
\dens(M_p) = \frac{1}{p}.
\end{equation}
\end{proposition}
\proof
Note that $M_p = p \otimes \mathbb{N}$.
Thus, according to Proposition~\ref{proposition.scaling},
we have that
\begin{align}
\dens(M_p)
=
\dens(p \otimes \mathbb{N})
=
\frac{1}{p} \dens(\mathbb{N})
=
\frac{1}{p}.
\end{align}
\endproof
\noindent
Consequently, we have, for instance,
$\dens(M_1) = \dens(\mathbb{N}) = 1$,
$\dens(M_2) = 1/2$,
$\dens(M_5) = 1/5$
etc.

\begin{figure}[t]
\centering
\subfigure[]{\epsfig{file=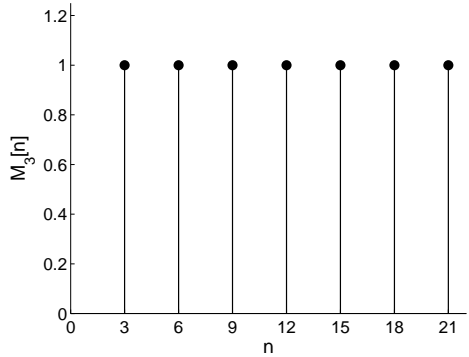,width=0.75\linewidth}}
\subfigure[]{\epsfig{file=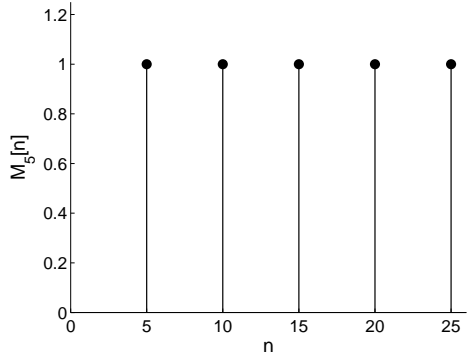,width=0.75\linewidth}}
\caption{Discrete-time signals $M_3[n]$ and $M_5[n]$
associated to the sets $M_3$ and $M_5$, respectively.}
\label{fig1}
\end{figure}

One can derive a physical interpretation for the
density of $M_p$.
Consider discrete-time signals $M_p[n]$ characterized by a sequence of discrete-time impulses
associated to the sets $M_p$.
The signals $M_p[n]$ are built according to a binary
function that returns one if $n$ is
a multiple of $p$, and zero otherwise.
For example, we can have $M_3[n]$ and $M_5[n]$, associated to
$M_3$ and $M_5$,
as depicted in Figure~\ref{fig1}.
In terms of signal analysis,
$\dens(M_p)$
has some correspondence to the
average value of $M_p[n]$.

\section{Density of Particular Sets}

In this section, some special sets have their density examined.
We focus our attention in three
notable sets:
(i) set of prime numbers,
(ii) Fibonacci sequence,
and
(iii)
set of square-free integers.

\subsection{Set of Prime Numbers}

Let $P =\{2, 3, 5, 7, 11, 13, 17, 19, 23, \ldots\}$
be the set of prime numbers.

\begin{proposition}
The prime numbers set is sparse.
\end{proposition}
\proof
Utilizing the definition of the prime zeta function,
$\zeta'(s)\triangleq\sum_{\text{$i:p_i$ is prime}}p_i^{-s}$,
we have that
$\dens(P) = \lim_{s\downarrow1}\frac{\zeta'(s)}{\zeta(s)}$.
Additionally observe that
$\zeta'(s)<\infty$ for $s>1$.
Taking in account that
\begin{equation}
\ln\zeta(s) = \sum_{k=1}^\infty \frac{\zeta'(ks)}{k},
\end{equation}
for $\epsilon>0$ we have that
\begin{align}
\ln\zeta(1+\epsilon)
&=
\sum_{k=1}^\infty \frac{\zeta'(k(1+\epsilon))}{k} \\
&=
\zeta'(1+\epsilon) + \sum_{k=2}^\infty \frac{\zeta'(k(1+\epsilon))}{k}.
\end{align}
Now consider the following inequalities:
\begin{align}
0\leq
\frac{\zeta'(1+\epsilon)}{\zeta(1+\epsilon)}
&=
\frac{\ln\zeta(1+\epsilon) - \sum_{k=2}^\infty \frac{\zeta'(k(1+\epsilon))}{k}}{\zeta(1+\epsilon)}
\\
&=
\frac{\ln\zeta(1+\epsilon)}{\zeta(1+\epsilon)} - \sum_{k=2}^\infty \frac{\frac{\zeta'(k(1+\epsilon))}{k}}{\zeta(1+\epsilon)}.
\\
&\leq
\frac{\ln\zeta(1+\epsilon)}{\zeta(1+\epsilon)}.
\end{align}
As $\epsilon\to0$, we have that $\zeta(1+\epsilon)\to\infty$, therefore
\begin{align}
\lim_{\epsilon\to0}\frac{\ln\zeta(1+\epsilon)}{\zeta(1+\epsilon)}
=
0,
\end{align}
because $\lim_{x\to\infty}\ln(x)/x = 0$.
Therefore,
we can apply the Squeeze Theorem once more, and find that
\begin{align}
\lim_{\epsilon\to0}\frac{\zeta'(1+\epsilon)}{\zeta(1+\epsilon)}
=
0.
\end{align}

\begin{figure}
\centering
\epsfig{file=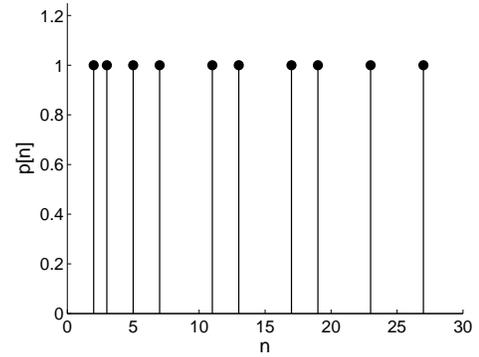,width=0.75\linewidth}
\caption{Discrete-time signal $P[n]$ corresponding to the set of prime numbers.}
\label{fig2}
\end{figure}

\endproof
Since the density is zero,
it indicates that the associated discrete-time signal $P[n]$, as shown in Figure~\ref{fig2},
has null average value.

\subsection{Fibonacci Sequence}

The Fibonacci sequence constitutes another interesting subject of investigation.
Fibonacci numbers are constructed according to the following recursive
equation
$u[n]=u[n-1]+u[n-2]$, $n>1$, for $u[0]=u[1]=1$, where $u[n]$ denotes the nth
Fibonacci number~\cite[p.160]{burton1998}.
This procedure results in the Fibonacci set
$F=\{1,2,3,5,8,13,21,34,55,\ldots\}$.

\begin{proposition}
The Fibonacci set is sparse.
\end{proposition}
\proof
It is known that the sum of the reciprocals of the Fibonacci numbers converges to
a constant (reciprocal Fibonacci constant),
whose value is approximately
$3.35988566\ldots$~\cite{jeannin1989}.
Consequently,
for
$s>1$,
we have that
\begin{align}
&\dens(F)
=
\lim_{s\downarrow1}
\frac{\sum_{n=2}^\infty\frac{1}{u[n]^s}}{\zeta(s)}
\\
&
<
\lim_{s\downarrow1}
\frac{\sum_{n=1}^\infty\frac{1}{u[n]}}{\zeta(s)}
=
\lim_{s\downarrow1}
\frac{3.35988566\ldots}{\zeta(s)}
=
0.
\end{align}
\endproof

\subsection{Set of Square-free Integers}

The set of square-free numbers can be defined as
$S\triangleq \{ n\in \mathbb{N} \ |\  |\mu(n)|=1 \}$,
where $\mu(\cdot)$ is the M\"obius function,
which is given by
\begin{align}
\mu(n)
=
\begin{cases}
1, & \text{if $n=1$}, \\
0, & \text{if $p^2|n$, for some prime number $p$}, \\
(-1)^r, & \text{if $n$ is the product of distinct}\\
 & \text{prime numbers.}
\end{cases}
\end{align}
Thus the first elements of $S$ are given by $1,2,3,5,6,7,10,11,13,14,15,\ldots$

\begin{proposition}
The density of the set of square-free integers is $\frac{1}{\zeta(2)}$.
\end{proposition}
\proof
It is known~\cite{schroeder1997} that $\sum_{n=1}^\infty \frac{|\mu(n)|}{n^s} = \frac{\zeta(s)}{\zeta(2s)}$,
thus it follows easily that:
\begin{align}
\dens(S)&=\lim_{s\downarrow1} \frac{\sum_{n\in S}\frac{1}{n^s}}{\zeta(s)}  \\
&=\lim_{s\downarrow1} \frac{\sum_{n=1}^\infty \frac{|\mu(n)|}{n^s}}{\zeta(s)} \\
&=\lim_{s\downarrow1} \frac{\frac{\zeta(s)}{\zeta(2s)}}{\zeta(s)} =\lim_{s\downarrow1} \frac{1}{\zeta(2s)}  \\
&=\frac{1}{\zeta(2)} =\frac{6}{\pi^2}.
\end{align}
\endproof

\section{Computational Results}

\begin{definition}
Let $A$ be a set of positive integers.
The truncated set $A^T$ is defined by
\begin{equation}
A^T=A \cap \{1, 2, 3, \ldots, T-1, T\},
\end{equation}
where $T$ is a positive integer.
\end{definition}
In other words, $A^T$ contains the elements of $A$ which are no greater than $T$.

The concept of truncated sets can furnish
computational approximations for the numerical value of
some densities.
Consider, for example,
the quantity
\begin{equation}
\label{definition.prob}
\asymdens'(A^T) \triangleq \frac{\|A^T\|}{\|\mathbb{N}^T\|}.
\end{equation}
This expression $\asymdens'(\cdot)$ can be interpreted in a frequentist way
as the ratio between
favorable cases and possible cases.
Clearly,
the asymptotic density can
be expressed in terms of $\asymdens'(\cdot)$:
\begin{equation}
\label{definicao.densidade.asymptotica.limite}
\asymdens(A)
=
\lim_{T\to\infty}\asymdens'(A^T).
\end{equation}

In a similar fashion,
we can consider a version of the discussed density
for truncated sets as defined below.
First,
observe that discussed density $\dens(\cdot)$ can be expressed as the following double limit
\begin{align}
\label{double.limit}
\dens(A) =
\lim_{s\downarrow 1}
\lim_{T\to\infty}
\frac{\sum_{n\in A^T} \frac{1}{n^s}}{\sum_{n=1}^T \frac{1}{n^s}}.
\end{align}
Restricted to the class of sets for which the above limits can have
their order interchanged,
we find that
\begin{align}
\dens(A) &=
\lim_{T\to\infty}
\lim_{s\downarrow 1}
\frac{\sum_{n\in A^T} \frac{1}{n^s}}{\sum_{n=1}^T \frac{1}{n^s}} \\
&=
\lim_{T\to\infty}
\left(
\frac{\sum_{n\in A^T} \frac{1}{n}}{\sum_{n=1}^T \frac{1}{n}}
\right).
\end{align}

\begin{definition}[Approximate Density]
\label{truncado}
The approximate density for a truncated set $A^T$ is given by
\begin{align}
\dens'(A^T) =
\frac{\sum_{n\in A^T} \frac{1}{n}}{\sum_{n=1}^T \frac{1}{n}}.
\end{align}
\end{definition}
Again,
whenever the limit order of~Equation~\ref{double.limit}
can be interchanged,
we have that
\begin{equation}
\dens(A) = \lim_{T\to\infty}\dens'(A^T).
\end{equation}
The quantity $\dens'(\cdot)$ furnishes
a computationally feasible way to investigate
the behavior of $\dens(\cdot)$.

In the following,
we obtain computational approximations for the asymptotic density
and the discussed density.
The approximate densities
are
numerically evaluated
as $T$ increases
in the range from 1 to 1000.
Now we investigate
(i)~arithmetic progressions,
(ii)~the set of prime numbers,
and
(iii)~the Fibonacci sequence.

\subsection{Arithmetic Progressions}
\begin{figure}
\centering
\epsfig{file=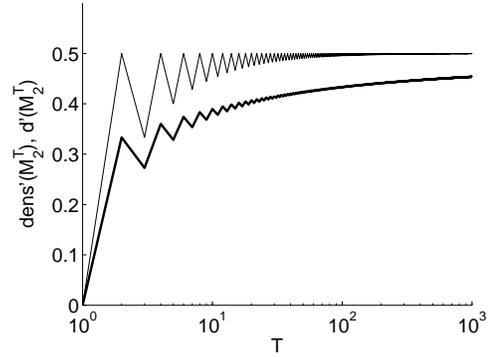,width=0.75\linewidth}
\caption{Computational result for the approximate density for the truncated prime set $M_2^T$,
according to (i) the function $\dens'(P^T)$ (bold curve),
and
(ii) the function $\asymdens'(P^T)$ (thin line).
Observe the convergence to $1/2$.}
\label{fig4}
\end{figure}

Considering an arithmetic progressions $M_q$,
we have that:
\begin{equation}
\label{relacao}
\asymdens(M_q)
=
\lim_{T\to\infty}\asymdens'(M_q^T)
=
\frac{1}{q},
\end{equation}
and
\begin{align}
\dens(M_q)
=
\lim_{T\to\infty}\dens'(M_q^T).
\end{align}
Both quantities limits converge to the same quantity $1/q$.
Intuitively,
we have that
the chance of ``selecting'' a multiple of $q$
among all integers is $1/q$.
This is exactly the density of $M_q$.
Figure~\ref{fig4} shows the result of computational calculations of
$\asymdens'(\cdot)$ and  $\dens'(\cdot)$ for $M_2^T$.

\subsection{Prime numbers sets}

Consider the truncated set of prime numbers
$P^T=\{p : \text{$p$ is a prime}, p<T)$.
One may easily compute both $\asymdens'(P^T)$ and $\dens'(P^T)$
for any finite $T$.

\begin{figure}[t]
\centering
\epsfig{file=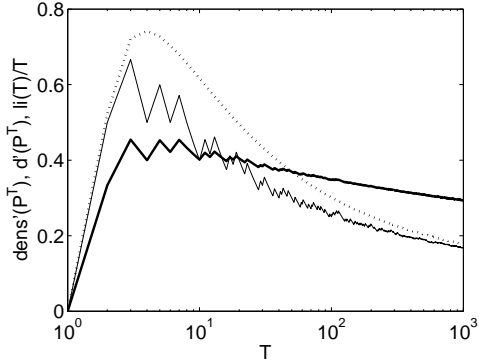,width=0.75\linewidth}
\caption{Computational result for the approximate density for the truncated prime set $P^T$,
according to (i)~the function $\dens'(P^T)$ (bold curve),
(ii)~the function $\asymdens'(P^T)$ (thin line),
and (iii)~the approximation based on the Prime Number Theorem (dotted line).}
\label{fig5}
\end{figure}

An alternative path for the estimation of $\asymdens'(P^T)=\frac{\pi(T)}{T}$,
where the function $\pi(x)$ represents the number of primes that do not exceed $x$,
is to utilize
the approximation for $\pi(T)$ given by
the Prime Number Theorem~\cite[p.336]{burton1998,riesel1985}:
\begin{equation}
\pi(T) \approx \li(T),
\end{equation}
where $\li(\cdot)$ denotes the logarithmic integral~\cite{abra1968}.
The curves displayed in Figure~\ref{fig5}
correspond to
the calculation of $\dens'(P^T)$, $\asymdens'(P^T)$ and $\li(T)/T$.
As expected, all curves decay to zero.

\subsection{Fibonacci Sequence}

\begin{figure}[t]
\centering
\epsfig{file=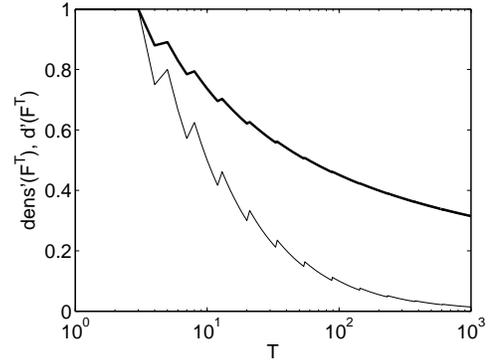,width=0.75\linewidth}
\caption{Density of the truncated Fibonacci set: $\dens'(F^T)$ (bold line) and $\asymdens'(F^T)$ (thin line).}
\label{fig7}
\end{figure}

Now consider the truncated Fibonacci set $F^T=\{1,2,3,5,8,13,21,\ldots,T\}$.
The result of calculating
\begin{equation}
\asymdens'(F^T)=
\frac{\|F^T\|}{\|\mathbb{N}^T\|}
\end{equation}
and
\begin{equation}
\dens'(F^T)=
\frac{\sum_{n\in F^T}\frac{1}{n}}{\sum_{n=1}^T\frac{1}{n}},
\end{equation}
are shown in Figure~\ref{fig7}.
Both curves tend to zero as $T$ grows.
This fact is expected since
we have already verified that
$\dens(F)=0$.
The small convergence rate of $\dens'(\cdot)$
is typical of problems that
deal with the harmonic series.

\section{Conclusion}

A density for infinite sets of integers was investigated.
It was shown that the suggested density shares several properties of usual probability.
A computational simulation of the proposed approximate density for
finite sets was addressed.
An open topic is the characterization of sets for which
the limiting value of the approximate density equals their density.

%
%
%
%

%\lipsum

%\section*{Acknowledgments}
%The first author would like to thank
%Dr.~P.~B.~Oliva,
%Department of Computer Science, Queen Mary, University of London,
%for many insightful commentaries and suggestions.

%

%

{\small
\bibliographystyle{IEEEtran}
\bibliography{ref}
}

\end{document}